\documentclass[conference]{IEEEtran}
\IEEEoverridecommandlockouts
\usepackage{cite}
\usepackage{amsmath,amssymb,amsfonts}
\usepackage{algorithmic}
\usepackage{graphicx}
\usepackage{textcomp}
\usepackage{xcolor}

\makeatletter
\def\ps@IEEEtitlepagestyle{ 
    \def\@oddfoot{\mycopyrightnotice}
}
\def\mycopyrightnotice{
    {\footnotesize 978--1--6654--1741--9/22/\$31.00~\copyright2022 IEEE\hfill}
}

\def\BibTeX{{\rm B\kern-.05em{\sc i\kern-.025em b}\kern-.08em
    T\kern-.1667em\lower.7ex\hbox{E}\kern-.125emX}}
\begin{document}

\title{Subspace Coding for Error Control and Interference Mitigation in Routing Solutions with Cooperative Destination Nodes\\
}

\author{\IEEEauthorblockN{Mohamed Amine Brahimi, Fatiha Merazka}
\IEEEauthorblockA{\textit{LISIC Lab.  Telecommunications Department,} \\
\textit{Electrical Engineering Faculty,}\\
\textit{USTHB University,}\\
16111, Algiers, Algeria\\
\{mbrahimi,fmerazka\}@usthb.dz}
}

\maketitle

\begin{abstract}
Random Linear Network Coding (RLNC) is a transmission scheme that opts for linear combinations of the transmitted packets at a subset of the intermediate nodes. This scheme is usually considered when Network Coding (NC) is desired over non-coherent networks. In order to integrate error correction in RLNC, subspace codes have been proposed. The codewords in those codes are unchanged under rank-preserving row operations, making them convenient for RLNC. In this paper, we investigate the use of those codes for the interference channel in a system model consisting of an array of SISO communication systems with cooperative destination nodes. This system model is deemed as a simplified model of a network with a routing based transmission scheme. Results have indicated that the use of subspace codes have allowed for better performance in terms of the decoding failure probability.
\end{abstract}


\section{Introduction}
Classical routing protocols are based on the store-and-forward mechanism \cite{Store}, where intermediate nodes are not allowed to alter the contents of their received packets. In fact, the payload of the transmitted packets in those systems is treated as a commodity flow and therefore packets are not to be combined or altered for the whole transmission session. In the seminal paper \cite{NC}, the authors introduced Network Coding (NC) as a paradigm for data transmission where packets may be combined during transmission to achieve higher throughput. The main contribution of \cite{NC} may be seen as the leap that authors made in how data is treated. In this regard, data is no longer considered as a commodity flow but rather as information flow. Using NC for practical applications was still challenging until Linear NC (LNC) was introduced in \cite{li_lnc_2003}. In this scheme, packets are represented as vectors from a finite field $\mathbb F_q$ and encoding operations are just linear tranformations of the original packets using coefficients from the underlying finite field. While this scheme was practically feasible, it still required apriori knowledge of network topology for the proper selection of the encoding coefficients. This limitation made LNC undesired in non-coherent situations where both the source and destination nodes are topology-oblivious. Random LNC (RLNC) \cite{RLNC1},\cite{RLNC2}was later proposed as a solution to this problem. In this setting, encoding coefficients are no longer deterministically selected but rather randomly chosen from a sufficiently large finite field. Note that the size of the finite field is important to minimize the probability of choosing an encoding coefficient that results in a linear dependency between the encoded packets \cite{fp}.

While NC schemes have been proven to outperform classical routing solutions in terms of throughput\cite{th1},\cite{th2}, security\cite{secure_algo},\cite{secure_subspace}... etc, they still face a set of challenges, of which error propagation is the crucial one. This is basically since packets are mixed in NC, which allows for errors to propagate from corrupt packets to valid ones.

In RLNC, subspace codes\cite{kotter_errorcoding_rnc_2008}, \cite{silva_rankmetric_2008} have been proposed for error control due to their properties that match those of how information is transmitted in RLNC. In those codes, codewords are seen as vector spaces from an ambient vector space over some finite field $\mathbb F_q$ where $q$ is a prime power. the correction capability of subspace codes is not affected by rank-preserving row operations and they provide error correction up to a certain number of row deletions and insertions, which makes them optimal for RLNC-based networks.

 The use of subspace codes is not generally considered when classical routing solutions are used instead of RLNC. In this paper, we try to investigate the effects of their deployment in those systems as compared to other classical error correction solutions. We prove that there exist situations where subspace codes may provide a degree of resilience against both network errors and crosstalk for classical routing systems with cooperative destination nodes.
 
 The remainder of this paper is organized as follows. In section II, we provide some background on subspace codes as well as the notation that is used throughout this paper. Section III will be dedicated to the system model, on which our work is based. In Section IV, we show how subspace codes can be used to provide both error control and interference mitigation for our system and Section V will constitute a general conclusion to the work done in this paper as well as a set of perspectives for eventual future work.

 \section{Preliminaries}
 A finite field with $q$ elements with $q=p_r^m$, where $p_r$ is a prime number and $m\in \mathbb N^*$, is denoted by $\mathbb F_q$. The $n$-dimensional vector space over $\mathbb F_q$ is denoted by $\mathbb F_q^n$ where $n\in \mathbb N^*$. Over this vector space, we define the projective space $\mathcal P(n)$, which is the  set of all subspaces of the vector space $\mathbb F_q^n$. The term "projective" comes from its geometrical meaning and uses in projective geometry, where it is also defined as the projective geometry of dimension $n-1$ over $\mathbb F_q$. A subset of this space with the criterion that all subspaces are of equal dimension is called a Grassmannian or Grassmann variety to link it to its geometrical properties. As for notation, a Grassmannian whose elements are the $k$-dimensional subspaces of  $\mathbb F_q^n$ is denoted by $\mathcal G(k,n)$. The relationship between the Grassmannians and the projective space on the vector space can be stated as follows,
 \begin{align}
\mathcal G(k,n)=\{  V| V\in \mathcal P(n),\; dim(V)=k, \; 0\leq k \leq n\}
\end{align}
In other words, the set of all Grassmannians over the vector space  $\mathbb F_q^n$ constitutes $\mathcal P(n)$.
We are intereseted in the aforementioned structures due to their importance in the definition of subspace codes.

A subspace code $\mathcal C$ can be defined as a non-empty subset of the projective space $\mathcal P(n)$. When $\mathcal C$ satisfies $\mathcal C \subseteq \mathcal G(k,n)$, with $0\leq k \leq n $ , we call $\mathcal C$  a Grassmannian code or a constant dimension code. Those codes are the ones that have attracted most of the attention in the litterature of subspace codes. 

An element $V\in \mathcal C $ will be represented by a matrix $M$ in the reduced row echelon form  and we write $V = \langle M \rangle $, where $\langle \cdot \rangle$ denote the row space.
  \section{System Model}
We consider an array of parallel single-input single-output (SISO) systems that are simplified into basic communication systems that comprise three elements : a source node, a destination node and a channel. The source node is the communicating entity that is the origin of the information transmitted on the system. The destination node is the communicating entity to which the source information is intended and the channel is the medium through which information is transmitted. While, in general, this system can be  full-duplex, we can safely consider the two opposite channels as being identical and therefore limit our analysis to only one channel ( direction : source $ \longrightarrow$ destination ). The source node has also the ability to perform encoding on the transmitted data to allow for error correction in the system. Similarly, the destination nodes are endowed with decoders to allow for the reverse operation.

The channels used in this system are all unit capacity channels, in which they can transmit one packet per time slot. Packets will be taken as $n$-dimensional vectors over a finite field $\mathbb F_q$. Fig .1. illustrates the system model.

We would like to allow for interference between the different channels of the system. Therefore, the set of channels form together an interference channel where all channels are affected by each other through electromagnetic coupling. In other words, the information transmitted in one channel will affect the information transmitted in the other $m-1$ channels. Note that this interference is usually referred to as crosstalk and therefore the  terms interference and crosstalk will be used interchangeably. In this paper, interference is modelled as the superposition of channel signals. In other words, if $M_i$ is the message or packet transmitted by a source $s_i$ to a destination $d_i$ with $i\in \{1,2,\cdots, m\}$, the received signal $R_i$ at $d_i$ will then take the form :

\begin{align}
    R_i=\sum_{j=1}^{m} \alpha_j M_j +Z_i
\end{align}
where $Z_i$ is the error vector that is assumed to be an element of $\mathbb F_q^n$. As for $\alpha_j$, if $j=i$, $\alpha_j=1$. Otherwise ( $j\neq i$), $\alpha_j$ is a discrete random variable that reflects the contribution of the corresponding signal in the output of the interference channel. A value of 0 means that the power of the interfering signal is too low and will not affect the channel output and therefore, it is safely neglected. However, a value of 1 will indicate that the signal power is high and therefore the corresponding vector will add up to the main signal. Concerning its distribution, $\forall j\in \{1,2,\cdots, m\} \;$ and $ j\neq i,\;\; \alpha_j$ is an independent Bernoulli random variable with the two probabilities $P(1)=p$, $P(0)=1-p$.

\begin{figure}[tbh!]
\centerline{\includegraphics[width=0.48\textwidth]{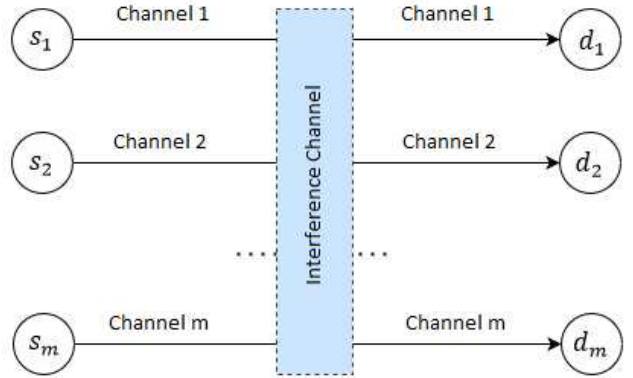}}
\caption{ System model}
\label{fig1}
\end{figure}
 The system model will later be modified into that in Fig.2 by introducing a new node $S$ considered as a main source that feeds information to all the other sources such that each source information is a distinct packet (from the other sources' fed packets). In this case, the rate of the source $S$  will be equal to $m$ packets/ time slot. Moreover, we will be treating interference as a form of uncontrolled RLNC with the random coefficients being taken from the binary field. To allow for data recovery, we will be assuming cooperation between the destination nodes to allow for decoding operations.
 \begin{figure}[tbh!]
\centerline{\includegraphics[width=0.5\textwidth]{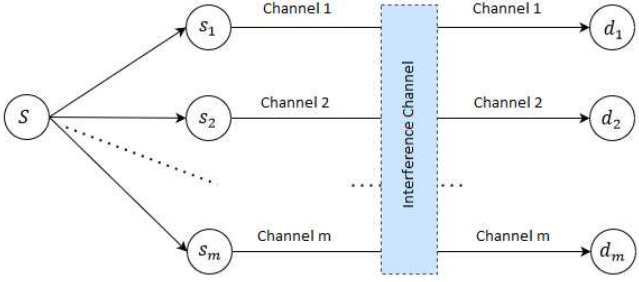}}
\caption{ Modified system model }
\label{fig2}
\end{figure}
 \section{Subspace Codes as a Solution for the Interference Channel}
As mentioned earlier in this paper, the codewords of a given subspace code are subspaces from  some vector space $\mathbb F_q^n$. This property has made subspace codes suitable for RLNC-based networks by treating data as a vector space instead of a set of independent packets.

Since the crosstalk problem that characterizes the interference channel can be seen as a superposition of information from different channels, one can make use of subspace codes as a solution to mitigate this problem while providing error correction capability for the system. Moreover, since routing systems are a set of independent unicast transmissions through a set of relays, one may simplify routing scenarios to an array of SISO systems (Fig. 1. or Fig .2.) by eliminating the intermediate relays, given that they do not affect information across the network. Therefore, a solution to the SISO system considered in this paper may be extended to networks that are based on a routing solution to convey data as long as the destination nodes are allowed to cooperate.
 In order to see the effects of subspace codes, we will be considering two scenarios: error-free transmission through the interference channel and transmission in the presence of errors.
 \subsection{Error-free Transmission}
In this scenario, equation (2) is simplified into 
\begin{align}
    R_i=\sum_{j=1}^{m} \alpha_j M_j
\end{align}
by eliminating the error vector.

To see the effects of susbspace codes, we try to consider three transmission schemes:
\begin{itemize}
    \item Routing: in this transmission, both system models (Fig.1 and Fig. 2) are acceptable and destination nodes are not cooperative ( they do not exchange any information).
     \item RLNC: in this transmission, we also consider that both system models (Fig.1 and Fig. 2) are allowed and destination nodes are cooperative ( they can exchange useful information to allow for RLNC decoding by assuming the interference signal as a form of uncontrolled RLNC encoding operation).
      \item Subspace coding: In this last scenario, only the system in Fig. 2. is acceptable. The reason behind this choice is the fact that while all the packets in this scenario are linearly independent, they are still vectors from the same codeword from a subspace code. The source $S$ will be thought of here as the source node with a subspace encoder.
\end{itemize}
In order to test the three scenarios, we will be using a set of 8 packets that are the vectors of the codeword in Fig.3. This codeword has been taken from the kk subspace code (16,256,16,8). In the first two scenarios, the vectors are used as being independent data packets. However, in the third scenario, they form the codeword in question. We have used the vectors of this codeword for the three scenarios to provide consistency for our experiment. Note that, in this system, $m=8$.
\begin{figure}[ht]
\label{fig3}
\centering

\centering
\begin{equation*}
\begin{bmatrix}

1& 0& 0& 0& 0& 0& 0& 0& 0& 0& 0& 0& 0& 0& 0& 1  \\
0& 1& 0& 0& 0& 0& 0& 0& 1& 0& 1& 1& 0& 0& 1& 1  \\
0& 0& 1& 0& 0& 0& 0& 0& 0& 1& 0& 1& 0& 1& 1& 1  \\
0& 0& 0& 1& 0& 0& 0& 0& 1& 0& 1& 0& 1& 1& 0& 1  \\
0& 0& 0& 0& 1& 0& 0& 0& 1& 1& 0& 0& 1& 0& 1& 1  \\
0& 0& 0& 0& 0& 1& 0& 0& 1& 1& 1& 1& 1& 1& 0& 1  \\
0& 0& 0& 0& 0& 0& 1& 0& 0& 1& 0& 0& 0& 0& 1& 1  \\
0& 0& 0& 0& 0& 0& 0& 1& 0& 0& 0& 1& 1& 1& 0& 1  \\
    
\end{bmatrix}
\end{equation*}

\caption{A codeword from the  (16,256,16,8) KK code.}
\end{figure}

The packets will be sent following the three aforementioned transmission scenarios while changing the value of $p$. We then evaluate the probability that the destination nodes will not be able to get the source message. This probability will be referred to as the decoding failure probability. While there is no decoding in the routing scenario, we will still be using this term for the sake of consistency. The results of this experiment are depicted in Fig .4. Note that the term `` Average decoding failure probability '' is used instead of `` Decoding failure probability '' in this latter figure because the experiment was repeated 1000 times and the average was taken as an estimation of the expected decoding failure probability.

From Fig. 4, we see that both RLNC and RLNC with subspace coding have allowed for the recovery of the intended source data packet for every destination node for all values of $p$. This means that since in those scenarios the destination nodes can cooperate to perform either RLNC or subspace decoding operation, the superposition of signals does not result in packet loss. As for the routing scenario, the decoding failure probability
depends on $p$. Since this latter parameter depicts the probability that interference occurs with the other signals. The results of Fig. 4 are consistent with what is expected. In fact, we see that as long as $p>0.3$, it is more likely that destination nodes will not be able to correctly receive their intended packets for the routing scenario.

\subsection{Transmission in the presence of errors}
The previous experiment will be repeated with the assumption that network errors occur on the network.

\begin{align}
    R_i=\sum_{j=1}^{m} \alpha_j M_j+Z_i
\end{align}

 Since the used packets and error vectors are vectors from the binary field, we can model errors as being bit-flips of the vector  $\sum_{j=1}^{m}  M_j$. In this case, the number of errors in a given vector will be seen as the number of non-zero entities in the error vector.  

The three previous transmission scenarios will also be adopted here. However, for this experiment, we will fix the value of $p$ to be $p=0.2$ while monitoring the performance of the system under the presence of errors. The rationale behind choosing a small value of $p$ i.e. 0.2, is based on the results of Fig.4. as a higher value of $p$ results in a higher average decoding failure probability per se, even in the absence of errors, which would  conceal the effects of errors on this latter.

Note that packets will be treated as the row vectors of a data matrix $R$ and errors are random bit-flips of the entries of the matrix $R^\prime$ whose row vectors are the vectors $R_i^\prime=R_i-Z_i$.

The results of this experiment are shown in Fig. 5. Routing alone is very sensitive to errors followed by RLNC which provides a slight improvement on the decoding failure probability. Subspace coding on the other side has resulted in a better performance where errors are less likely to affect the decoding process as long as they are under 40.

 \begin{figure}[tbh!]
\centerline{\includegraphics[width=0.55\textwidth]{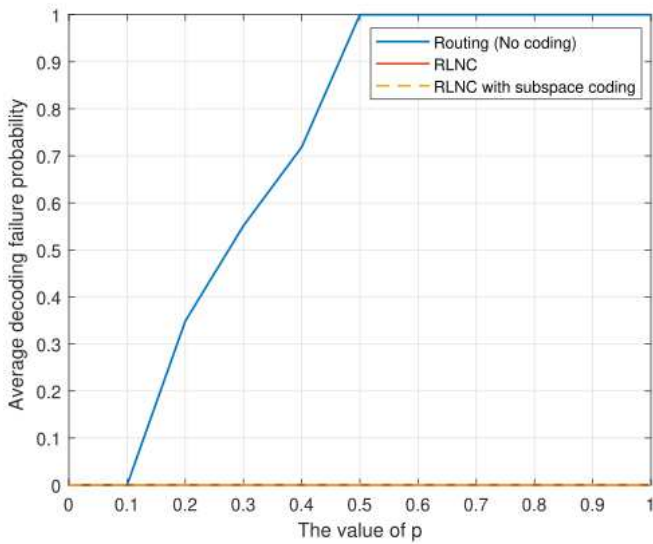}}
\caption{ Decoding failure probability Vs. the value of $p$ for the three transmission scenarios : Routing, RLNC, and RLNC with subspace coding in an error-free environment across the interference channel. }
\label{fig4}
\end{figure}

 \begin{figure}[tbh!]
\centerline{\includegraphics[width=0.53\textwidth]{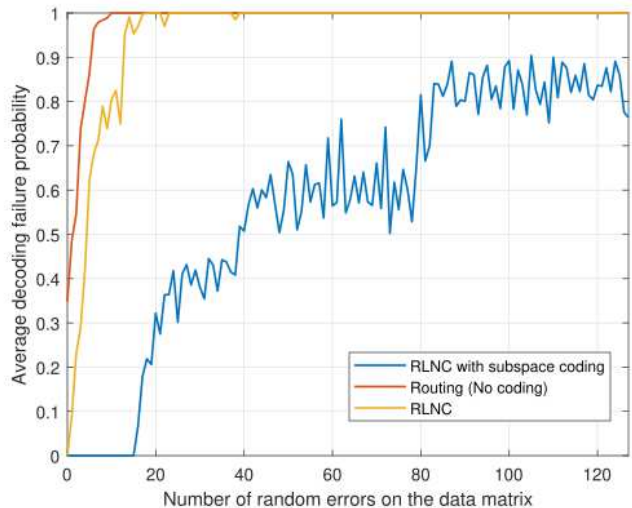}}
\caption{ Decoding failure probability Vs. the number of random errors in the data matrix for the three transmission scenarios : Routing, RLNC, and RLNC with subspace coding  across the interference channel with $p=0.2$. }
\label{fig5}
\end{figure}

\section{Conclusion}
In this paper, we have investigated the use of subspace codes as a solution for the interference channel with the absence as well as the presence of errors for a system model consisting of an array of SISO communication systems with cooperative destination nodes. Interference may be deemed as a form of uncontrolled RLNC with coefficients taken from the binary field following a Bernoulli distribution. In an error-free environment, as long as the destination nodes are cooperative and able to perform RLNC decoding operations, interference will not result in data loss. However, in the presence of errors, cooperation amongst destination nodes will not result in a significant change in the error performance of the system. In this case, the use of subspace codes will be useful as the results of our experiments suggest. This comes from the fact that subspace codes are designed to correct errors affecting subspaces and by treating data in our system as a vector space, subspace codes will provide error correction as well as interference mitigation.

As for our future work, we will focus on the effects of other code families on mitigating the effects of interference as well as providing error control solutions for the system such as rank-metric codes.

\end{document}